\documentclass{aa}  

\usepackage{graphicx}
\usepackage{txfonts}
\usepackage{color}

\begin{document}

   \title{X-ray long-term variations in the low-luminosity AGN NGC\,835 and its circumnuclear emission}
   \titlerunning{Obscuration in the LLAGN NGC\,835}


   \author{
 O. Gonz\'alez-Mart\'in\inst{1}\thanks{Tenure track at IRAF (\email{o.gonzalez@crya.unam.es})}
\and
L. Hern\'andez-Garc\'ia\inst{2}
\and
J. Masegosa\inst{2} 
\and  
I. M\'arquez\inst{2} 
\and 
J.M. Rodr\'iguez-Espinosa\inst{3,4}
\and \\
J.~A. Acosta-Pulido\inst{3,4} 
\and
A. Alonso-Herrero\inst{5,6}
\and 
D. Dultzin\inst{7}
\and 
D. Esparza Arredondo\inst{7}
}

   \institute{
Instituto de Radioastronom\'ia y Astrof\'isica (IRAF-UNAM), 3-72 (Xangari), 8701, Morelia, Mexico \and 
Instituto de Astrof\'isica de Andaluc\'ia, CSIC, Glorieta de la Astronom\'ia s/n 18008, Granada, Spain \and  
Instituto de Astrof\'isica de Canarias (IAC), C/V\'ia L\'actea, s/n, E-38205 La Laguna, Spain \and 
Departamento de Astrof\'isica, Universidad de La Laguna (ULL), E-38205 La Laguna, Spain \and 
Instituto de F\'isica de Cantabria, CSIC-UC, E-39005 Santander, Spain \and
Visiting professor, Department of Physics and Astronomy, University of Texas at San Antonio, San Antonio, TX 78249, USA \and 
Instituto de Astronom\'ia, Universidad Nacional Aut\'onoma de M\'exico, Apartado Postal 70-264, 04510 M\'exico DF, Mexico}
   \date{Received ??; accepted ??}

 
  \abstract
   { Obscured active galactic nuclei (AGNs) are thought to be very common in the Universe. Observations and surveys have shown that the number of sources increases for near galaxies and at the low-luminosity regime (the so-called LLAGNs). Furthermore, many AGNs show changes in their obscuration properties at X-rays that may suggest a configuration of clouds very close to the accretion disk. However, these variations could also be due to changes in the intrinsic continuum of the source. It is therefore important to study nearby AGN to better understand the locus and distribution of clouds in the neighbourhood of the nucleus.}
   {We aim to study the nuclear obscuration of LLAGN NGC\,835 and its extended emission using mid-infrared observations.  }
   {We present sub-arcsecond-resolution mid-infrared 11.5$\rm{\mu m}$ imaging of the LLAGN galaxy NGC\,835 obtained with the instrument CanariCam in the Gran Telescopio CANARIAS (GTC), archival \emph{Spitzer}/IRS spectroscopy, and archival \emph{Chandra} data observed in 2000, 2008, and 2013. }
   {The GTC/CanariCam 11.5$\rm{\mu m}$ image reveals faint extended emission out to $\rm{\sim}$6 arcsec. We obtained a nuclear flux of $\rm{F(11.5\mu m) \sim18~mJy,}$ whereas the extended emission accounts for 90\% of the total flux within the 6 arcsec. This
means that the low angular resolution ($\rm{\sim}$4 arcsec) IRS spectrum is dominated by this extended emission and not by the AGN. This is clearly seen in the \emph{Spitzer}/IRS spectrum, which resembles that of star-forming galaxies. Although the extended soft X-ray emission shows some resemblance with that of the mid-infrared, the knots seen at X-rays are mostly located in the inner side of this mid-infrared emission. The nuclear X-ray spectrum of the source has undergone a spectral change between 2000/2008 and 2013. We argue that this variation is most probably due to changes in the hydrogen column density from $\rm{\sim 8\times 10^{23}cm^{-2}}$ to $\rm{\sim 3\times 10^{23}cm^{-2}}$.  NGC\,835
therefore is one of the few LLAGN, together with NGC\,1052, in which changes in the absorber can be claimed.}
   {}


   \maketitle
%

\section{Introduction}

The emission in active galactic nuclei (AGNs) is powered by accretion onto a supermassive black hole (SMBH). AGNs are traditionally divided into two main classes based on the presence (type 1) or absence (type 2) of broad permitted lines (FWHM$\rm{>}$2000 km $\rm{s^{-1}}$) in the optical spectrum. The so-called unification model (UM) proposes that both types of AGNs are essentially the same objects viewed at different angles \citep{Antonucci93,Urry95}. An optically thick dusty torus surrounding the central source would then be responsible for blocking the region where these broad emission lines are produced in type 2 AGNs. Thus, the obscuration is the key in the UM of AGNs. 

Low-ionisation nuclear emission-line regions (LINERs), first classified by \citet{Heckman80}, are the dominant population of AGNs in the local Universe \citep{Ho97}, and all of them are within the class of low-luminosity AGN (LLAGN, i.e. $\rm{L_{bol}<10^{42}erg/s}$). However, they remain one of the most captivating subsets of nuclear classes because the main physical mechanism powering LLAGN is still unknown. The nature of LINERs was initially sustained in their optical spectrum, which can be reproduced with a variety of different physical processes \citep[e.g. photoionisation from hot stars, non-thermal photoionisation, shocks, post-main sequence stars, or AGN,][]{Dopita95,Heckman80,Ferland83,Veilleux87,Cid-Fernandes10, Stasinska08,Singh13}. Today we know from multi-wavelength information
that around 75-90\% of LINERs show evidence of AGNs \citep{Gonzalez-Martin06,Gonzalez-Martin09A,Gonzalez-Martin15,Dudik09,Younes11,Asmus11,Mason12,Maoz05,Hernandez-Garcia13,Hernandez-Garcia14}. 

Some authors have argued that strong obscuration is responsible for the differences to more luminous AGNs \citep[e.g.][]{Dudik09,Gonzalez-Martin09B}. Using the ratio between the luminosity of the [OIII]$\rm{\lambda 5007\AA{}}$ emission line and the intrinsic hard (2-10 keV) X-ray luminosity, L([OIII])/$\rm{L_{X}(2-10~keV)}$, as a tracer of Compton thickness (i.e. $\rm{N_{H}>1.5\times10^{24}~cm^{-2}}$), \citet{Gonzalez-Martin09B} found that up to 53\% of the LINERs in their sample are Compton-thick candidates. This percentage is twice as high as that reported for type 2 Seyferts \citep{Maiolino98,Bassani99,Panessa06,Cappi06}. \citet{Dudik09} studied the emission lines in 67 high-resolution \emph{Spitzer}/IRS spectra of LINERs and found that the central power source in a high percentage of LINERs is highly obscured at optical frequencies, consistent with the X-ray results. 

In this paper we report a new case of absorption variations in a LLAGN that we found through the joint analysis of long-term X-ray variations and mid-infrared emission. The dust that absorbs the shorter wavelength emission reradiates in the mid-infrared and correspondingly produces a substantial fraction of the bolometric flux of the object. Thus, mid-infrared observations can give good evidence of the dust heated by the AGN in the very centre \citep[e.g.][]{Ramos-Almeida11}, together with information on the star formation when present \citep[e.g.][]{Esquej14,Alonso-Herrero14}. We note that high angular resolution is needed to distinguish nuclear \citep[torus emission located at the inner $\rm{\sim}$10\,pc,][]{Tristram09,Burtscher13,Asmus14} from extranuclear emission. This can currently only be achieved with 10m class ground-based telescopes. 

We have conducted observations of a small sample of LLAGN with the instrument CanariCam \citep{Telesco03,Packham05} on the 10.4m Gran Telescopio CANARIAS (GTC) in La Palma. The observations include images with the Si-5 filter at 11.5$\rm{\mu m}$ of 18 LLAGN. Part of the data is already available and has been used to compare large-scale \emph{Spitzer} fluxes with the nuclear fluxes in \citet{Gonzalez-Martin15}. The full sample will be published in a forthcoming paper. 

To contribute to understanding the obscuration in LLAGN, here we report the X-ray multi-epoch analysis together with the mid-infrared analysis of NGC\,835, a spiral galaxy classified as an SAB that belongs to the Hickson compact Group 16 \citep[HCG\,16, ][]{Hickson82}. Its nucleus has been classified as a LINER \citep[][]{Martinez10} or as a Seyfert \citep{Veron10}. No broad lines were found in the optical spectrum, classifying it as a type 2 AGN \citep{Jones09}. Throughout this paper, we use a distance for NGC\,835 of 34 Mpc\footnote{The distance of 34 Mpc corresponds to a redshift of z=0.007939 (using ${\rm H_{0}=70}$ km/s/Mpc).}, the redshift-independent measurements taken from the NASA extragalactic database (NED\footnote{http://ned.ipac.caltech.edu}). Note that 1 arcsec corresponds to $\rm{\sim}$165 pc. 

\citet{Gallagher08} studied the mid-infrared emission in 12 nearby HCGs using \emph{Spitzer} nuclear photometry, finding a mid-infrared flux at 8$\rm{\mu m}$ of 136mJy for this source. They argued that this flux is most probably powered by star formation, although they claimed that mid-infrared spectra were needed to answer
this unambiguously. \citet{Bitsakis14} computed an infrared 8-1000$\rm{\mu m}$ luminosity of $\rm{log(L_{IR}/L_\odot)=10.02}$ for NGC\,835, using far-infrared and sub-millimetre  \emph{Herschel} observations. Nevertheless, this object has never been observed at mid-infrared with high spatial resolution.

\begin{figure*}[!t]
\begin{center}
\includegraphics[width=1.\columnwidth]{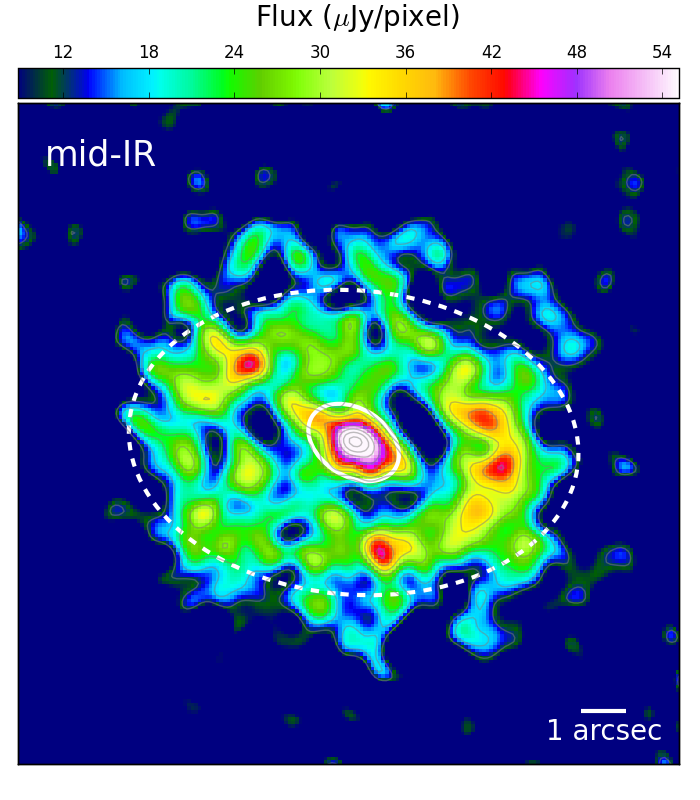}
\includegraphics[width=1.\columnwidth]{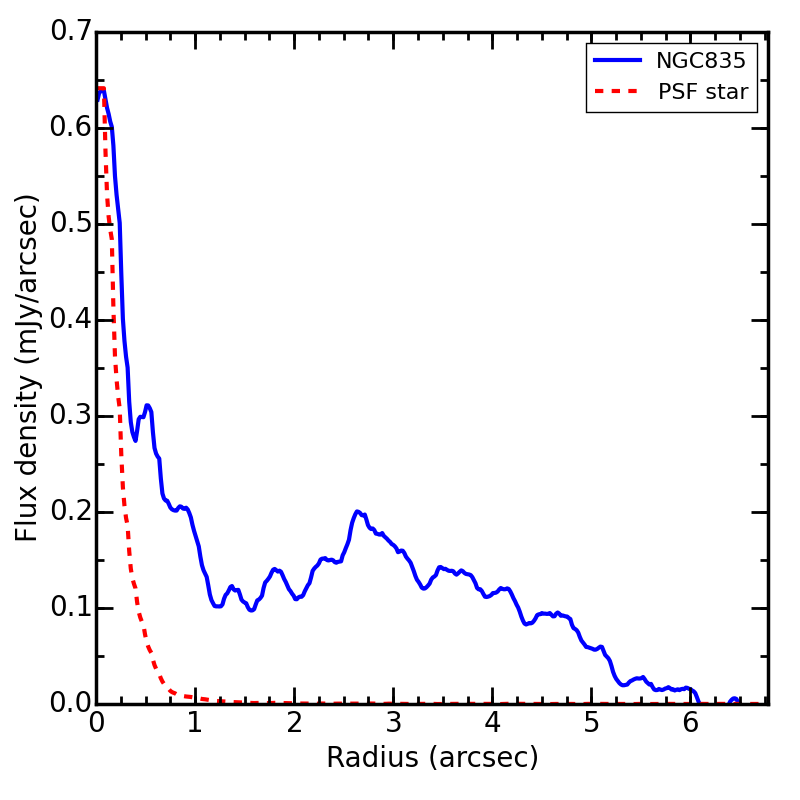}
\caption{(Left): Flux-calibrated mid-infrared image of Si5 filter at 11.5$\rm{\mu m}$ taken with CanariCam/GTC. This image is centred at the nucleus of NGC\,835. The contour level with the lowest value is at 5.5$\rm{\sigma}$. The continuous and dashed (white) ellipses show the best fit to the inner and outer ring, respectively (see text). North is up,  east is left. (Right): Radial profile of NGC\,835 centred at the nuclear position (blue solid line) compared to the PSF radial profile (dashed red line). }
\label{fig:midinfrared}
\end{center}
\end{figure*}

At X-rays, \citet{Turner01} confirmed the AGN nature of NGC\,835 using the available \emph{XMM}-Newton spectrum. Snapshot \emph{Chandra} data were first presented for this object by \citet{Dudik05}, who found that the circumnuclear emission at X-rays was made of multiple hard off-nuclear point-like sources of comparable brightness to the nuclear source \citep[see also][]{Gonzalez-Martin06}. They obtained an intrinsic hard X-ray luminosity of $\rm{L(2-10 keV)=7\times 10^{39}}$ erg/s and a very low accretion rate ($\rm{L_{bol}/L_{Edd}=2\times 10^{-6}}$). \citet{Gonzalez-Martin09A} and \citet{Gonzalez-Martin09B} later included this object in a large sample of LLAGN, finding that the source was possibly Compton thick (i.e. hydrogen column density of $\rm{N_{H}>1.5\times 10^{24}cm^{-2}}$). Recently, \citet{OSullivan14} presented new \emph{Chandra} observations of NGC\,835 (together with the full system HCG\,16). They found variations in the hard X-ray band compared to previous \emph{Chandra} observations, which they attributed most probably to changes in the accretion disk. In this paper we simultaneous fitted the \emph{Chandra} observations with the mid-infrared nuclear analysis and come to a different conclusion. Furthermore, we report a complete analysis of the circumnuclear emission of the source. Mid-infrared CanariCam/GTC data are presented in Sect. \ref{sec:mid-infrared}, and X-ray \emph{Chandra} data are shown in Sect. \ref{sec:X-ray}. Finally, a full discussion of the results is included in Sect. \ref{sec:discussion}, and the main conclusions are presented in Sect. \ref{sec:conclusions}. 

\section{Mid-infrared CanariCam/GTC data}\label{sec:mid-infrared}

We present the first sub-arcsecond resolution mid-infrared imaging of NGC\,835. This nucleus was observed with Canaricam/GTC on 23 September 2014. Images were taken using the Si5 filter (at 11.5$\rm{\mu m}$ with an effective width of 0.9$\rm{\mu m}$) in two separate observing blocks with a total on-source time of 993\,s \footnote{Two observing blocks were observed to produce individual observing blocks shorter than one hour, according to the GTC policy.}. 

CanariCam uses a Raytheon 320$\rm{\times}$240 Si:As detector that covers a field of view (FOV) of 26$\rm{\times}$19 arcsec on the sky with a pixel scale of 0.0798 arcsec. The standard mid-infrared chopping-nodding technique was used to remove the time-variable sky background, the thermal emission from the telescope, and the detector 1/f noise, where f is the frequency of the noise component. The employed chopping and nodding throws were 10 arcsec, with chop and nod position angles of 180$\rm{^{o}}$ and 0$\rm{^{o}}$, respectively.

These observations are part of proprietary data of a sample of low-luminosity and Compton-thick LINERs observed with CanariCam/GTC (proposal ID GTC10-14A, P.I. Gonz\'alez-Mart\'in). The observations of the entire sample are not yet complete and will be the focus of a subsequent publication \citep[see][for preliminary results]{Masegosa13}. 

Images in the same filter of the point spread function (PSF) standard star HD\,11353 (on-source exposure time of 66\,s) were obtained immediately after the science target to accurately sample the image quality and allow for flux calibration of the target observation. The angular resolution of the observations is 0.24 arcsec (39.6 pc), as computed from the full width at half-maximum (FWHM) of the observed PSF standard star.

Each observing block was processed using the pipeline RedCan \citep[][]{Gonzalez-Martin13}, which is able to produce flux--calibrated imaging and wavelength-- and flux--calibrated spectra for CanariCam/GTC and T-ReCS/Gemini low-resolution data. The combination of the two observing blocks was made after flux-calibration with Python routines. 

Figure \ref{fig:midinfrared} (left) shows the final flux-calibrated 11.5$\mu m$ image obtained with CanariCam. The nucleus corresponds to the brightest source in the image. The total flux of the nuclear source (computed using aperture photometry centred at the peak of the nuclear source and with a radius of 1 arcsec) is 20.0 mJy. However, this nuclear region is not point-like. Figure \ref{fig:midinfrared} (right) shows the radial profile of NGC\,835 (blue continuous line) and the PSF profile of the standard star scaled to the peak of the emission in NGC\,835 (red dashed line). Even the central 1 arcsec shows a contribution of extended emission. We fitted the inner 1 arcsec to a 2D Gaussian, showing that it is elongated along $\rm{i=58^{o}}$ (north to east) with a FWHM of 0.87 and 0.60 arcsec, along and perpendicular to the elongation, respectively (the extension of this 2D Gaussian is shown as a continuous white ellipse in Fig.\ref{fig:midinfrared}, left). This elongated structure extends up to a radius of $\rm{\sim}$2 arcsec from the nuclear source. We used the scaled profile of the standard star to derive that the nuclear (point-like) source flux in NGC\,835 is $\rm{\sim}$18.4 mJy within 1 arcsec. Thus, the extended emission contributes 8\% to the total flux in the inner 1 arcsec. The external contribution increases up to $\rm{\sim}$92\% when we consider the inner 6 arcsec of the source (total flux of 246.8 mJy). This extended emission, clearly seen up to 6 arcsec from the nucleus of NGC\,835 (see Fig. \ref{fig:midinfrared}), has a ring-like morphology. We fitted the extended emission to an ellipsoid to characterise its extension\footnote{The ellipsoid used for the extended emission describes its extension and not the peak of the emission.} , obtaining a position angle consistent with zero, a horizontal major axis of $\rm{\sim}$4.1 arcsec ($\rm{\sim}$680 pc), and a vertical minor axis of $\rm{\sim}$2.8 arcsec ($\rm{\sim}$460 pc, shown as a dashed white ellipse in Fig.\ref{fig:midinfrared}, left). This ring-like structure is not equally distributed, showing two bright spots. The first one is located at a position angle of $\rm{PA=58^{o}}$, consistent with the inner elongated direction mentioned before. The second and more extended spot is shown with orientation angles in the range of $\rm{PA=185-290^{o}}$, with three smaller spots within this region.

\begin{table}
\caption{X-ray observations and  $\rm{\chi_{r}^{2}}$ for the individual spectral fits.}
\label{tab:X-raysdata}      
\centering                                      
\begin{tabular}{r c c c}          
\hline\hline                        
 ObsID & Date & Net Exposure (ksec) & $\rm{\chi_{r}^{2}}$ (indiv. fit) \\ \hline
923      & 11/16/2000 & 12.6 & 0.98 \\
10394  & 11/23/2008 & 13.8 & 0.97 \\
15181  &  7/16/2013 &  49.5 & 0.97 \\
15666  &  7/18/2013 &  29.7 & 0.96 \\
15667  &  7/21/2013 &  58.3 & 0.98 \\
\hline                                             
\end{tabular}
\end{table}

\section{X-ray \emph{Chandra} data}\label{sec:X-ray}

We have focused the X-ray analysis presented in this paper on the public \emph{Chandra} data because their superb resolution is the best to be compared with our mid-infrared image. We refer to \citet{Gonzalez-Martin06} and \citet{Gonzalez-Martin09A} for a full discussion on the X-ray data, both with \emph{Chandra} and \emph{XMM-Newton} satellites. Furthermore, for a complete analysis of the \emph{Chandra} data of Hickson 16 we refer to \citet{OSullivan14}. 

Five archival \emph{Chandra} observations are available. Table \ref{tab:X-raysdata} shows the details of these observations. All the data were processed following standard procedures within the CXC Chandra Interactive Analysis of Observations (CIAO, v4.6.3\footnote{http://cxc.harvard.edu/ciao/}) package and analysed using Xspec software (v12.8.2\footnote{http://heasarc.gsfc.nasa.gov/xanadu/xspec/}). Level-2 event files were extracted by using the {\sc acis-process-events} task. We first cleaned the data from background flares (i.e., periods of high background) that could affect our analysis. To clean them we used the {\sc lc\_clean.sl} task, which removes periods of anomalously low (or high) count rates from light curves from source-free background regions of the CCD. In the subsequent analysis we study the X-ray spectra and images. 

\begin{figure*}[!t]
\begin{center}
\includegraphics[width=1.\columnwidth]{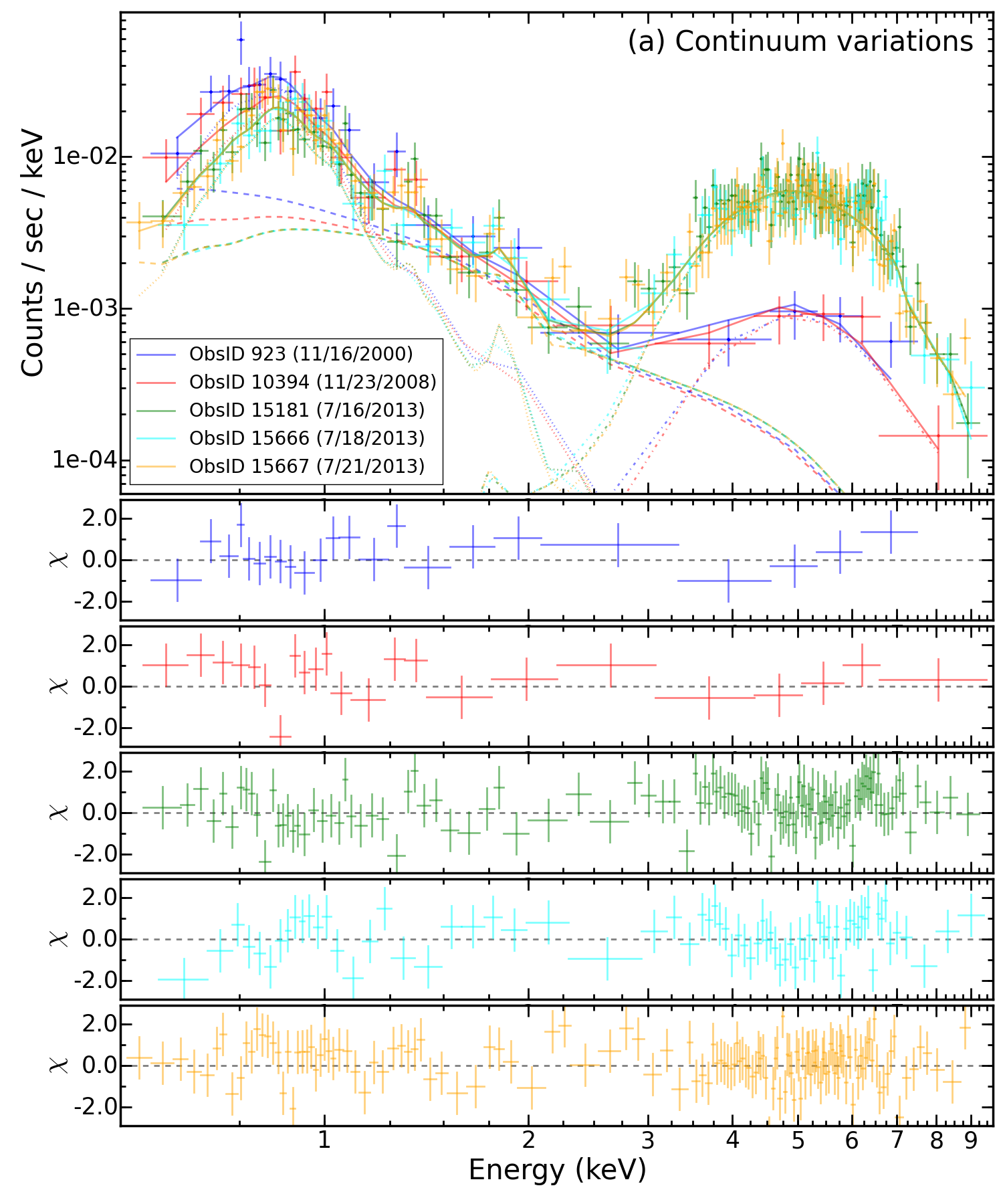}
\includegraphics[width=1.\columnwidth]{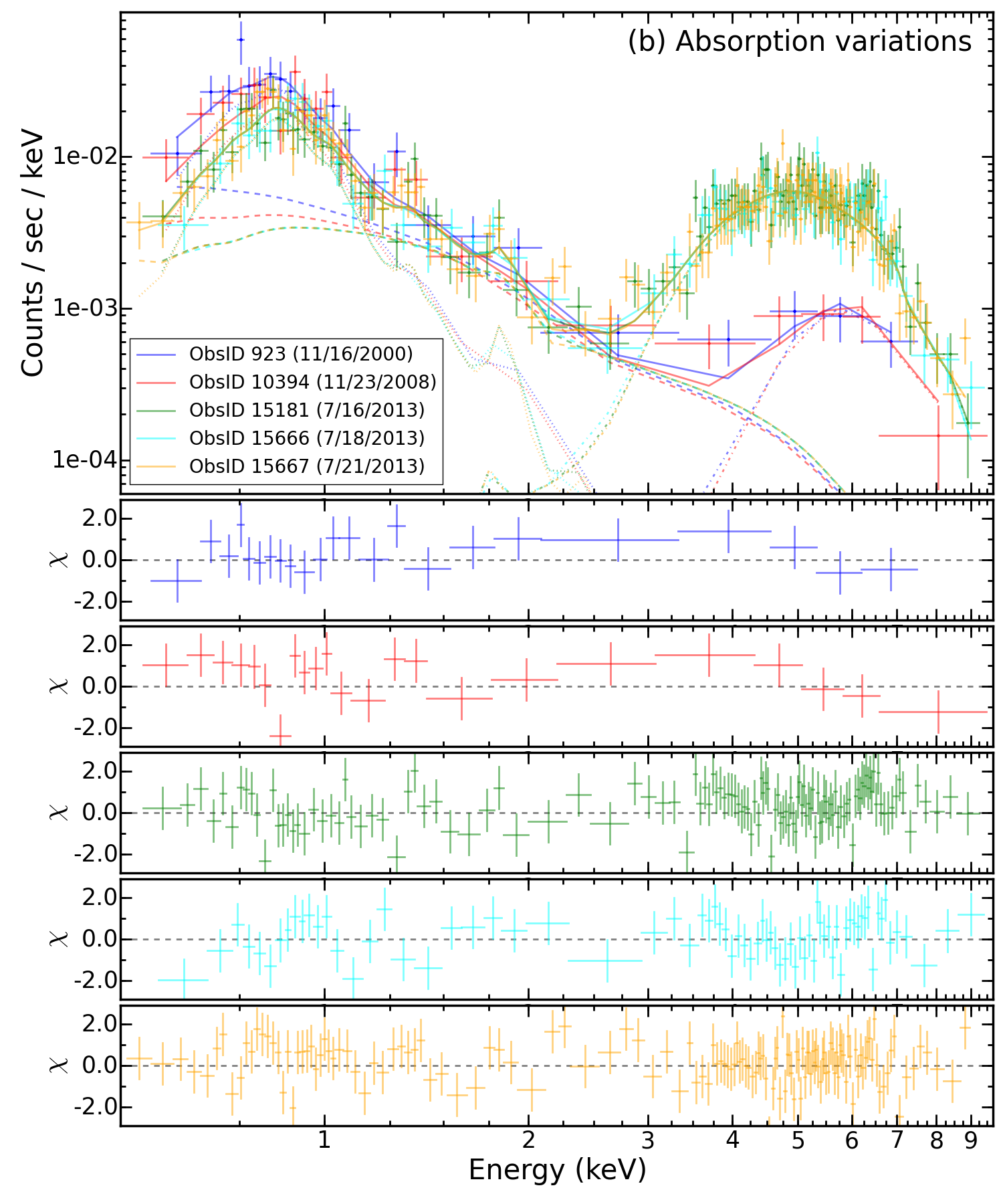}
\caption{Simultaneous spectral fit to the five X-ray observations, including changes in (a) the intrinsic continuum flux (left) and in (b) the absorber in the line of sight (right). The top panel for each of them shows the spectral fitting for all the observations with different colours (see legend). The thermal, scattering, and intrinsic continuum components are also shown with dotted, dashed, and dot-dashed lines, respectively. The five panels below show the residuals for each of the observations separately.  }
\label{fig:Xrayfitting}
\end{center}
\end{figure*}

\begin{table*}
\caption{Results of the X-ray spectral fitting. X-ray luminosities in units of $\rm{10^{40} erg/s}$. Normalisations ($N_{1}$,  $N_{2}$, and  $N_{3}$) are in units of photons keV$\rm{^{-1}}$cm$\rm{^{-2}}$s$\rm{^{-1}}$ at 1 keV. Parameters marked with asterisks are set to be tied for all the observations. Including the FeK$\rm{\alpha}$ slightly improved the final statistic of the fit with a  $\rm{\chi_{r}^{2}\simeq 0.90}$ for both continuum and absorption variations (see text). }           
\label{tab:X-rays}      
\centering                                      
\begin{tabular}{r c c | c c}          
\hline\hline                        
   & \multicolumn{2}{c}{(a) Continuum variations} & \multicolumn{2}{c}{(b) Absorption variations}  \\   
                                                                        & 923/10394 & 15181/15666/15667 & 923/10394 & 15181/15666/15667 \\ \hline
  $\rm{\chi_{r}^{2}}$   \dots                 &                \multicolumn{2}{c|}{0.94}                              &         \multicolumn{2}{c}{0.95}                            \\
 T (keV)                \dots                           &   $\rm{0.61\pm0.03^{*}}$                         &                                                                         &  $\rm{0.61\pm0.03^{*}}$                          &                                    \\
 $N_{1}$                \dots                           &   $\rm{(1.2\pm0.1)\times 10^{-5*}}$ &                                                                      &  $\rm{(1.2\pm0.1)\times 10^{-5*}}$    &                                   \\
 $N_{2}$        \dots                                   &   $\rm{(7.9\pm0.9)\times 10^{-6*}}$ &                                                                      &  $\rm{(8.2\pm0.9)\times 10^{-6*}}$    &                                    \\
 $N_{H} (10^{22} cm^{-2})$ \dots        &    $\rm{32.3\pm1.6^{*}}$                        &                                                                          &  $\rm{89.7\pm12.3}$                          &   $\rm{32.0\pm1.7}$   \\
 $N_{3}$                \dots                           &   $\rm{(2.1\pm0.9)\times 10^{-4}}$ &$\rm{(1.16\pm0.08)\times 10^{-3}}$ &  $\rm{(1.14\pm0.08)\times10^{-3*}}$ &                                    \\
 EW(FeK$\rm{\alpha}$)  (eV)  \dots  &   $\rm{800\pm320}$                        &      $\rm{140\pm60}$                           &    $\rm{310\pm170}$                        &      $\rm{110\pm60}$    \\
 $L(0.5-2 keV)$ (obs.)  \dots           &    $\rm{0.67\pm0.02}$                 &     $\rm{0.67\pm0.20}$                          &  $\rm{0.67\pm0.02}$                      &    $\rm{0.67\pm0.02}$     \\
 $L(2-10 keV)$ (obs.)   \dots           &    $\rm{2.4\pm0.3}$                 &     $\rm{20\pm8}$                              &  $\rm{3.8\pm0.4}$                     &    $\rm{11.9\pm0.3}$      \\
 $L(0.5-2 keV)$ (intr.) \dots           &    $\rm{7.3\pm0.2}$                          &     $\rm{7.3\pm0.2}$                        &  $\rm{36.2\pm1.1}$                              &     $\rm{36.2\pm1.1}$      \\
 $L(2-10 keV)$ (intr.)  \dots           &    $\rm{8.6\pm1.1}$                          &     $\rm{46\pm18}$                                   &  $\rm{45.2\pm4.7}$                              &      $\rm{45.2\pm4.7}$      \\
\hline                                             
\end{tabular}
\end{table*}

\subsection{Spectroscopy}\label{sec:XraySpec}

Nuclear spectra for each observation were extracted from a circular region centred at the position of the brightest hard X-ray point source coincident with the coordinates provided by the NASA Extragalactic Database (NED\footnote{http://ned.ipac.caltech.edu}) for NGC\,835. The radius of these circular regions was fixed to 7 arcsec for all the observations. The background regions were extracted using a circular region at the same position for all the observations at $\rm{\sim}$40 arcsec toward the north of the source with a radius of 17 arcsec. 

We used the {\sc dmextract} task to extract the spectra of the source and the background regions. The response matrix file (RMF) and ancillary reference file (ARF) were generated for each source region using the {\sc mkacisrmf} and {\sc mkwarf} tasks, respectively. Before background subtraction, the spectra were binned to have a minimum of ten counts per spectral bin, to be able to use the $\rm{\chi^{2}}$-statistics, using the {\sc grppha} task included in FTOOLS\footnote{http://heasarc.gsfc.nasa.gov/ftools/ftools\_menu.html}.

We then performed a multi-epoch spectral fitting of the five epochs included in this analysis. We used several baseline models to determine long-term variations and to establish the main driver of these variations. This multi-epoch spectral fitting has been
fully described and tested in \citet{Hernandez-Garcia13} and \citet{Hernandez-Garcia14} for a sample of LINERs and in \citet{Hernandez-Garcia15A} for type 2 Seyferts. For clarity, we summarise the method here. First, the spectra are fitted individually with six models, comprising various combinations of absorbed thermal contributions and power
laws. We used the MEKAL model as a representation of a thermal contribution and a power law as a representation of a non-thermal contribution. We used f-test and $\rm{\chi^2}$ to define the best model for each individual observation. The best-fit model for all the data was defined as the more complex model needed to represent any of the individual spectra. The best model that fits all the observations is a combination of a soft thermal model (MEKAL) plus an unabsorbed power law to fit the soft emission (i.e. below $\rm{\sim}$2 keV) and an absorbed power law to fit the hard X-ray emission \citep[i.e. ME2PL, see][]{Hernandez-Garcia13}:

\begin{equation}
F(E) = Thermal(T,N_{1}) + N_{2} E^{-\gamma} + e^{-N_{H}\sigma (E)} N_{3} E^{-\gamma}
,\end{equation}

\noindent where $N_{1}$, $N_{2}$, and $N_{3}$ are the normalisations for each component, $\gamma$ is the slope of the power laws (both slopes tied to vary together to the same value), and $N_{H}$ is the hydrogen column density of the obscured component. The first power law represents a scattered component associated with the intrinsic continuum and mostly affects the soft energies (i.e. $\rm{<}$ 2 keV). The second power-law represents the intrinsic continuum of the AGN and is affected by the obscuration in our line of sight. We note that Galactic absorption was included in the model and was fixed to the predicted value using the {\sc nh} tool within FTOOLS \citep{Dickey90, Kalberla05}. The individual  $\rm{\chi_{r}^{2}}$ are reported in Table \ref{tab:X-raysdata}.

We have used this baseline model to produce a simultaneous fit for the full set of spectra. To do this, the spectral index was fixed to $\rm{\gamma=1.9}$ \citep[standard power-law index for AGNs, see e.g.][]{Panessa06,Bianchi09,Brightman11} to better constrain other parameters. This baseline model is unable to produce a good fit for all the data together without allowing variations of some parameters ($\rm{\chi_{r}^{2}= 2.3}$). Observations with ObsIDs 923 and 10394 (taken in 2000 and 2008, respectively) show fully consistent X-ray spectra without variations. The same occurs for the spectra with ObsIDs 15181, 15666, and 15667 (taken in 2013). However, an increase in the observed flux above $\rm{\sim}$2 keV is seen for the observations in 2013 compared to those in 2000/2008, as noted by \citet{OSullivan14}. For this reason, we separated the parameters in the two blocks; the first block includes ObsIDs 923 and 10394, and the second block includes ObsIDs 15181, 15666, and 15667. We did not find variations in the soft emission ($\rm{<}$2 keV).

We performed the first simultaneous fit to try to understand the main driver (i.e. parameter) governing this variation. We found that to describe the full set of data with the same model, changes in the normalisation ($\rm{\chi_{r}^{2}= 0.94}$) or the hydrogen column density $\rm{N_{H}}$ ($\rm{\chi_{r}^{2}= 0.95}$) associated with the hard X-ray power law are equally acceptable (statistically). Figure \ref{fig:Xrayfitting} shows the best
fit and residuals for all the observations, and Table \ref{tab:X-rays} shows the resulting parameters in the two scenarios. 

If the variations were due to changes in the normalisation of the hard X-ray power law, an increase of a factor of $\rm{\sim}$5 in the intrinsic continuum would be needed to explain the observed variations. Thus, the intrinsic continuum luminosity of the AGN changes from $\rm{\sim9\times 10^{40} erg/s}$ to $\rm{\sim5\times 10^{41} erg/s}$ in a five-year interval, from 2000/2008 to 2013. On the other hand, if changes in the obscuration were responsible for the observed variations, the source would have undergone a decrease of the $\rm{N_{H}}$, from $\rm{\sim 9\times10^{23} cm^{-2}}$ in 2000/2008 to $\rm{\sim3\times 10^{23} cm^{-2}}$ in 2013 (see Table \ref{tab:X-rays}). 

The spectrum can be described statistically without need of emission lines. However, the FeK$\rm{\alpha}$ line has been detected before in ObsIDs 15181, 15666, and 15667 by \citet{OSullivan14}. This is partially due to the technique of grouping the data. While the $\rm{\chi^{2}}$-statistic is required to select the best-fit model, it hides our observations from narrow features such as the FeK$\rm{\alpha}$ line \citep[see][for more details]{Guainazzi05}. We included the FeK$\rm{\alpha}$ line in our model, fixing the width of the line to 100 eV \citep[typical value for narrow lines in AGNs,][]{Jimenez-Bailon05}. This slightly improved the statistic ($\rm{\chi_{r}^{2}=0.90}$ for both models, i.e. changes in the continuum and in the absorber). The $\rm{EW(FeK\alpha)}$ for each scenario is recorded in Table \ref{tab:X-rays}.

The low count-rate of the extended emission prevents us from studying it separately. However, the hard-X-ray (i.e. $\rm{>2keV}$) component is confined to the central $\rm{\sim}$1 arcsec of the source with an almost negligible contribution below 2 keV. This, together with the results of the multi-epoch fitting, suggests that the former extended component is fully consistent with being
fixed and is associated with the soft emission detected spectroscopically. 

\begin{figure}[!t]
\begin{center}
\includegraphics[width=1.\columnwidth]{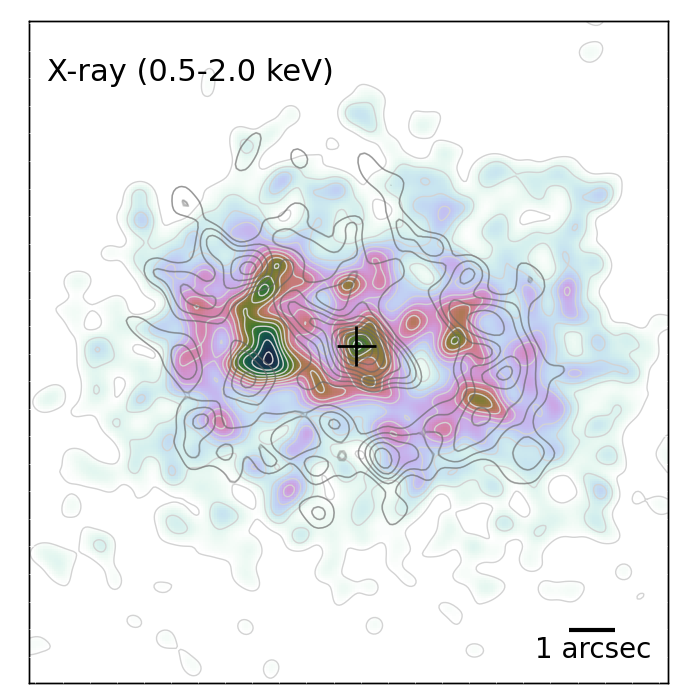}
\caption{Soft X-ray (0.5-2.0 keV) smoothed image of NGC\,835. Grey contours show the mid-infrared image taken with CanariCam/GTC as in Fig.~\ref{fig:midinfrared}. The large cross indicates the centroid of the hard X-ray point-like source used to align soft X-rays and mid-infrared images. North is left, east is down.}
\label{fig:Xrayimage}
\end{center}
\end{figure}

\subsection{Imaging}

The hard X-ray range morphology is point-like with no extended emission. We refer to \citet{OSullivan14} for more details on the hard X-ray morphology. Furthermore, it shows an increase in the observed flux in 2013 (ObsIDs 15181, 15666, and 15667) compared to previous observations in 2000 (ObsID 923) and 2008 (ObsID 10394), that were reported spectroscopically previously. The soft X-ray emission  is not variable in our five observations (see Sect. \ref{sec:XraySpec}), which we used to stack all the observations to improve the signal-to-noise ratio. This stacking was made using the {\sc reproject\_obs} task within the CIAO utilities. The output of this task is an event file with all the events reprojected on the same coordinate grid. We then produced smoothed 0.5-2.0 keV images using the {\sc dmcopy} (to chose the 0.5-2.0 keV energy range) and {\sc csmooth} (to use adaptive smoothing to enhance the extended emission) tasks. This image was constructed by subdividing the default pixel of \emph{Chandra} (0.492 arcsec/pix) to reach 0.05 arcsec/pix. The resulting 0.5-2.0 keV image is shown in Fig. \ref{fig:Xrayimage}. 

The nuclear component in the 0.5-2.0 keV X-ray image is not the brightest source in the image. This is consistent with the X-ray spectral analysis, which shows that the intrinsic continuum of the AGN is bright above $\rm{\sim}$ 2 keV. The high obscuration of the source (in the two possible scenarios, see the previous section and Table~\ref{tab:X-rays}) depresses the AGN continuum below these energies and makes it undetectable. The soft X-ray emission is quite extended with emission up to $\rm{\sim}$6 arcsec from the centre. The emission is elongated in the north-south direction (i.e. left to right). Compared to the mid-infrared CanariCam image (grey contours overlaid in Fig. \ref{fig:Xrayimage}), the 0.5-2~keV emission is more extended than that shown in the mid-infrared. We aligned the X-ray and mid-infrared images that showed the centroid of the hard X-rays (shown as a large cross in Fig. \ref{fig:Xrayimage}) to the centroid of the inner source at mid-infrared frequencies. This might be an indicator of different origins for each wavelength, but it might also be caused by the lower sensitivity of the mid-infrared image compared with the X-ray data. The morphology of the 0.5-2.0 keV X-ray image also shows a complex structure. A close comparison of the X-ray and mid-infrared images shows that these structures are not well correlated. Instead, the soft X-ray emission tends to be placed in the gaps of the mid-infrared emission. This is clearly seen in the brightest structure of the 0.5-2.0 keV X-ray image ($\rm{\sim}$2 arcsec toward the left of the figure), which is placed in the mid-infrared gap outside the inner nuclear emission of NGC\,835. The ring-like morphology observed toward the right of the figure at the 0.5-2.0 keV image also clearly lies before the ring-like structure observed in the mid-infrared image. It seems here that the X-ray ring is surrounded by the mid-infrared emission. We discuss the possible explanation of this displacement between soft X-rays and mid-infrared emission in Sect. \ref{sec:discussion}.

\section{Discussion}\label{sec:discussion}

We have analysed the nuclear and circumnuclear emission of the LINER source NGC\,835. We used proprietary mid-infrared continuum image (CanariCam/GTC) and public X-ray  data (\emph{Chandra}). Here we discuss the implications of these results in the context of the circumnuclear (Sect. \ref{sec:circumnuclear}) and nuclear (Sect. \ref{sec:nuclear}) emission. 

\begin{figure}[!t]
\begin{center}
\includegraphics[width=1.\columnwidth]{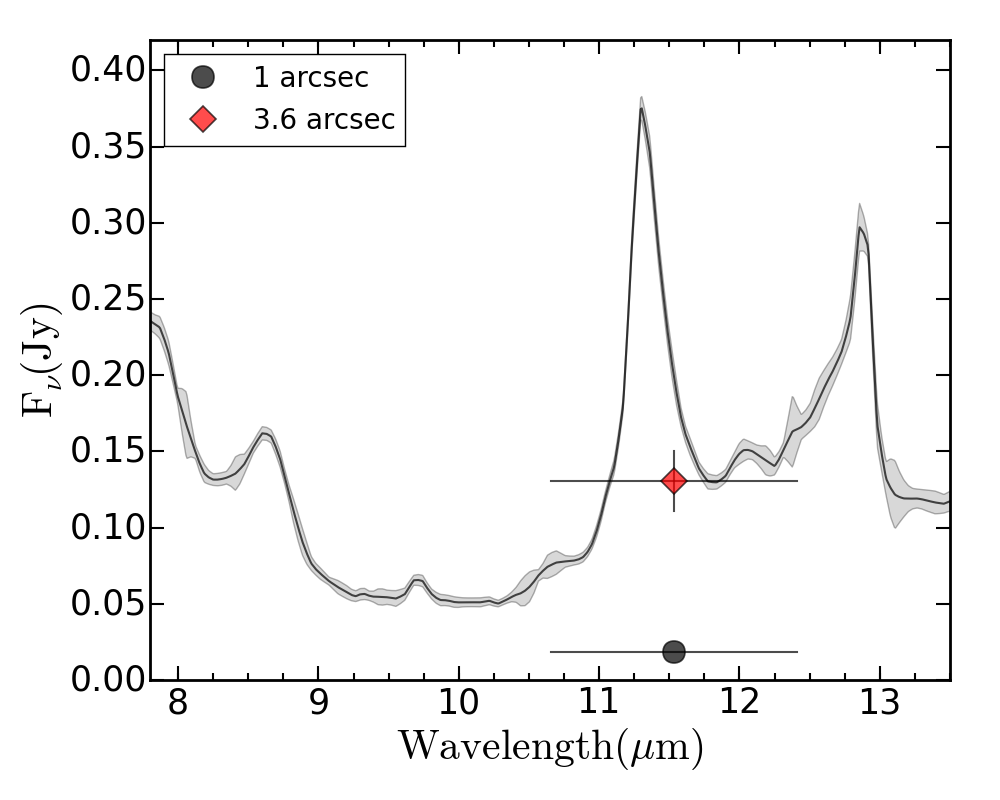}
\caption{\emph{Spitzer}/IRS spectrum of NGC\,835. Black circle and red diamond show the PSF-nuclear flux and the total flux using the slit width of \emph{Spitzer}/IRS (3.6 arcsec) using CanariCam data. The X-axis error bars in the photometric measurements denote the width of the filters. }
\label{fig:ComparisonSpitzer}
\end{center}
\end{figure}

\subsection{Circumnuclear emission}\label{sec:circumnuclear}

The mid-infrared image at 11.5$\rm{\mu m}$ reveals an extended emission composed by knots within a ring-like structure at $\sim$6 arcsec from the nuclear region (which corresponds to a radius of $\rm{\sim 1kpc}$). This extended emission accounts for $\rm{\sim}$90\% of the total emission within the central $\sim$6 arcsec. 

We downloaded the mid-infrared low spectral resolution spectrum of NGC\,835\footnote{The short-low spectrum obtained with \emph{Spitzer}/IRS has a slit width of 3.6 arcsec, much larger than the spatial resolution of CanariCam/GTC data, and covers a range between 5-30$\rm{\mu m}$.} from the Cornell Atlas of \emph{Spitzer}/IRS Sources (CASSIS\footnote{http://cassis.astro.cornell.edu}) to compare it with our mid-infrared fluxes (see Fig. \ref{fig:ComparisonSpitzer}). The most intense features are the 11.3$\rm{\mu m}$ PAH  band and the [Ne II] emission line at 12.8$\rm{\mu m}$, both tracers of star formation in the central 3.6 arcsec (i.e. $\rm{\sim 800}$ pc). The shape of the mid-infrared spectrum as observed by \emph{Spitzer} is very similar to the average spectrum of star-forming galaxies and very different from AGN-dominated LINERs \citep[see][]{Gonzalez-Martin15}, indicating that it is fully dominated by the circumnuclear emission. The \emph{Spitzer}/IRS flux at 11.5$\rm{\mu m}$ is 172$\rm{\pm}$26 mJy, consistent with the 11.5$\rm{\mu m}$ flux (131$\rm{\pm}$ 20 mJy) using as aperture the slit width used for the \emph{Spitzer/IRS} spectrum (i.e. 3.6 arcsec). However, the nuclear ($\rm{\sim 54}$pc) 11.5$\rm{\mu m}$ flux observed with CanariCam/GTC shows a lower flux of 18$\rm{\pm}$ 3 mJy. We interpret this as an extranuclear origin of the 11.3$\rm{\mu m}$ PAH feature, most probably associated with the extended emission seen in our ground-based CanariCam data. This was previously observed in some type 2 Seyferts, where the spectrum of the extended emission was dominated by the 11.3$\rm{\mu m}$ PAH feature at these wavelengths \citep{Alonso-Herrero14}. Indeed, the extended emission seen in the CanariCam image is fairly consistent with the 7.4 arcsec extended emission found in the \emph{Spitzer}/IRS spectrum by the CASSIS analysis\footnote{see http://cassis.astro.cornell.edu/atlas/cgi/radec.py?ra=32.35245\&\\dec=-10.135872\&radius=20}. Therefore, the extended mid-infrared emission seen with our mid-infrared CanariCam data is most probably associated with star-forming regions. 

Although the general morphology of the X-ray image matches that of the mid-infrared data, the emission knots seen in the two images are not in the same position. These knots seem to be displaced, with the mid-infrared structure being outside of the X-ray emission. Assuming that the ground-based mid-infrared image is a good tracer of star-formation, we therefore have to rule out a star formation origin for the extended soft X-ray emission. We note, however, that this extended soft X-ray emission seems to be intimately linked to that of the star-forming regions (because the soft X-rays are in the inner edges of the star-forming regions seen at mid-infrared wavelengths). This is also the case for instance of IRAS\,19252-7245. In this case the soft X-rays are produced in a galactic wind driven by the starburst \citep{Jia12}. 

The possible origins for the soft X-ray extended emission are photoionisation by the AGN or shocks. The optical line emission of some LINERs is known to be dominated by shock excitation \citep{Dopita95,Alonso-Herrero00,Ho08}, although some contribution of photoionisation is also invoked \citep{Masegosa11}. \citet{Ramos-Almeida14} tested this hypothesis by comparing the soft X-ray emission with the [O\,III] 5007 \AA{} emission for Mrk\,1066. The morphologies did not match, which rules out the photoionisation origin as proposed in other type 2 Seyferts \citep{Bianchi06,Gomez-Guijarro15}. However, to determine whether this soft X-ray emission is produced by shocks and/or photoionisation by the AGN, new observations\footnote{An H$\rm{\alpha}$ image is available observed with the CTIO 1.5m telescope obtained within the SINGG sample \citep[][]{Meurer06}. However, the low spatial resolution of this image does not allow us to compare it with our X-ray and mid-infrared images.} are required (for instance optical emission lines such as [O\,III] 5007 \AA{} emission to trace photoionisation).

\subsection{Nuclear emission}\label{sec:nuclear}

The nucleus of NGC\,835 was reclassified by \citet{Gonzalez-Martin06} as a LINER source using optical spectra. It shows signs of AGN activity at X-rays with a bright point-like source at hard energies \citep[i.e. $\rm{>}$2 keV,][]{Gonzalez-Martin09A}. The spectrum at hard energies is dominated by an absorbed power law (see Fig.~\ref{fig:Xrayfitting}), also suggestive of the presence of an obscured AGN \citep[i.e. type 2 AGN,][]{Gonzalez-Martin09A,OSullivan14}. 

\citet{OSullivan14} compared the best fit obtained for the two earliest \emph{Chandra} observations (2000 and 2008) with the three latest observations in 2013, finding that the nucleus of NGC\,835 is variable in the hard X-ray band on timescales from months to years. They suggested that this variation might be due to changes in the intrinsic continuum of the source. However, different scenarios were not tested. We have found that flux variations of the intrinsic continuum and absorption variations can explain the observed variability pattern equally well (see Sect. \ref{sec:XraySpec}). It is clear that the variability must have influence on scales of months or years because we do not detect it in the three observations taken within a few days in 2013. 

\citet{Gonzalez-Martin15} showed that one of the best ways to study the Compton-thick nature of the sources is by using the correlation between the X-ray and mid-infrared luminosities \citep{Gandhi09}, which nicely extends to low-luminosity AGN \citep{Mason12,Asmus14}. All the Compton-thick candidates included in \citet{Gonzalez-Martin09B} fall into this correlation only when their X-ray luminosities are corrected for Compton thickness. Thus, we can use this relation to study if the source has gone through a Compton-thick phase in 2000/2008. It is worth noting that by using the mid-infrared emission to determine the cause of the X-ray variations, we assume that the mid-infrared emission has not varied during this time. The large distance between the mid-infrared emission and the central source together with the fact that the total mid-infrared emission is the sum of all the individual clouds that form the torus suggests that mid-infrared variations is not relevant.

The nuclear mid-infrared flux inferred from our CanariCam/GTC observations at 11.5$\rm{\mu m}$ is 18.4 mJy. This implies a mid-infrared luminosity of $\rm{\lambda L_{\lambda} (11.5\mu m) = 7.3\times 10^{41}}$ erg/s at the distance of NGC\,835. \citet{Gandhi09} found a linear correlation between the mid-infrared and the hard (2-10 keV) X-ray luminosities as follows:  $\rm{ log (L_X) = 0.88\,log (L_{MIR}) + 4.75}$. According to this relation, the X-ray luminosity of the source should be $\rm{L_{X}\sim 3.9\times 10^{41}}$ erg/s. In the scenario in which the intrinsic continuum has changed, the luminosity has increased from $\rm{L_{X}\sim 0.86\times 10^{41}}$ erg/s to $\rm{L_{X}\sim 4.6\times 10^{41}}$ erg/s. The second epoch (i.e. 2013) is fairly consistent with the expected value. However, if the source was Compton-thick in the first epoch, we would expect the real luminosity to be
between 10 to 70 times higher than is observed (i.e. $\rm{L_{X}\sim [9-60]\times 10^{41}}$ erg/s), which is much higher than predicted by this correlation. On the other hand, in the scenario in which the absorption has changed, the X-ray intrinsic luminosity is $\rm{L_{X}\sim 4.5\times 10^{41}}$ erg/s, consistent with the mid-infrared to X-ray luminosity correlation. 

Therefore, changes in the absorption are preferred to explain the X-ray variability of this source. In this scenario the line is consistent with the same EW ($\rm{EW(FeK\alpha)=310\pm170}$ eV and $\rm{EW(FeK\alpha)=110\pm60}$ eV, respectively), as expected since the ratio between the intrinsic continuum and the reflection component has not changed. However, in this case, the source remains mildly obscured but Compton-thin in the two epochs. We
note that the use of mid-infrared information can help in other cases (as it does for this object) to understand the possible mechanism driving the X-ray variability in AGN. These variations are common among type 1.8 and 1.9 Seyferts \citep[e.g.][]{Risaliti07,Puccetti07,Risaliti11} even though they are not so common among type 2 Seyferts \citep[only four out of the 25 type 2 Seyferts analysed by][]{Hernandez-Garcia15A} and LINERs \citep[only in NGC\,1052 among the 17 LINERs analysed, ][]{Hernandez-Garcia13,Hernandez-Garcia14}. Thus, this object corresponds to a small group of LLAGN showing variations in the absorber. Sometimes these variations are observed fast enough to argue that they must be located very close to the accretion disc \citep[e.g. NGC\,1365, ][]{Risaliti11}. Unfortunately, the $\rm{N_H}$ variations were not observed close enough (in time) to constrain the locus of the absorber in NGC\,835. Monitoring campaigns would be needed to study the absorbers along the line of sight for this source. 

\section{Conclusions}\label{sec:conclusions}

We presented here a high angular resolution (0.3 arcsec) mid-infrared image obtained with CanariCam on the GTC of the LINER source NGC\,835 together with public X-ray \emph{Chandra} data. Here we summarise the main findings of this work. 
 
   \begin{itemize}
      \item The extended emission seen at mid-infrared for NGC\,835 shows a ring-like morphology within $\rm{\sim 1 kpc}$ radius (6 arcsec). The nuclear emission only accounts for 8\% of the total emission within the inner $\rm{\sim 1 kpc}$. This extended emission is most probably associated with star formation. However, the soft X-ray morphology does not match that of the mid-infrared. Soft X-rays are located at the inner side of the star-forming knots. The two possibilities are shocks and/or photoionisation by the AGN. More observations are needed to confirm the origin of the extended soft X-ray emission for this source.
      \item The nucleus of NGC\,835 is clearly dominated by the AGN emission at both X-rays and mid-infrared frequencies. Furthermore, the observed emission of this nucleus has undergone variations between the first two epochs of X-ray observations (in 2000 and 2008) and the latest three (in 2013). We found that these variations are most likely due to changes in the absorber according to the X-ray to mid-infrared luminosity relation expected for AGN. We propose that the combination of multi-epoch X-ray spectral fitting and nuclear mid-infrared luminosity could be very useful for understanding the variability processes in other AGN. 
   \end{itemize}

\begin{acknowledgements}
We thank to the anonymous referee for the fruitful comments and suggestions that have helped us to improve the final manuscript. OGM thanks C. Carrasco-Gonzalez for his help with the physical interpretation of the results. This scientific publication is based on observations made with the Gran Telescopio Canarias (GTC), installed at the Spanish Observatorio del Roque de los Muchachos of the Instituto de Astrof\'isica de Canarias on the island of La Palma.This research has been supported by the UNAM and the Spanish Ministry of Economy and Competitiveness (MINECO) under the grant (project refs. AYA2013-42227-P, AYA 2012-39168-C03-01, and AYA 2010-15169) and by La Junta de Andaluc\'ia (TIC 114). LHG acknowledges financial support from the Ministerio de Econom\'{i}a y Competitividad through the Spanish grant FPI BES-2011-043319. AAH acknowledges support from from MINECO AYA2012-31447 grant, which is partly funded by the FEDER program. DD acknowledges support from grant 107313 from PAPIIT, UNAM.  
\end{acknowledgements}


\end{document}